\documentclass[%
 reprint,
 amsmath,amssymb,
 aps,
floatfix,
]{revtex4-2}

\usepackage{graphicx}
\usepackage{dcolumn}
\usepackage{bm}

\usepackage{amsmath}
\usepackage{graphicx}
\usepackage{booktabs}
\usepackage{tabularx}
\usepackage[dvipsnames]{xcolor}
\usepackage{dcolumn}
\usepackage{bm}
\usepackage{xspace} 
\usepackage{rotating}
\usepackage{array}

\newcommand{\FNAL}{\affiliation{Fermi National Accelerator Laboratory, Batavia, IL 60510, USA}}

\usepackage{fancyhdr}
\begin{document}

\rhead{Contribution to the 25th International Workshop on Neutrinos from Accelerators}
\title{Updates and Lessons Learned from NuMI Beamline at Fermilab \\
\small (Contribution to the 25th International Workshop on Neutrinos from Accelerators) }

\author{D. A. A.\, Wickremasinghe}
\email{athula@fnal.gov}\FNAL

\author{K.\, Yonehara}\FNAL
\date{\today}
\begin{abstract}
The Neutrinos at the Main Injector (NuMI) beamline at Fermilab generates an intense muon neutrino beam for the NOvA (NuMI Off-axis $\nu_e$ Appearance) long-baseline neutrino experiment. Over the years, the NuMI beamline has been pivotal in advancing neutrino physics, providing invaluable data and insights. This proceeding paper discusses updates and the lessons learned from recent experiences during the beam operations, maintenance, and monitoring of the NuMI beamline. Key topics include the optimization of beam performance and challenges in maintaining beamline stability. The paper aims to share best practices and provide a road-map for future beamline projects, including the Long-Baseline Neutrino Facility (LBNF).
\end{abstract}

\maketitle
\section{Introduction}
The Neutrinos at the Main Injector (NuMI) beamline at Fermilab \cite{Adamson:2015dkw, NumiTechNote,Barenboim:2002mf} has been instrumental in advancing our understanding of neutrino physics, primarily through its role in generating an intense muon neutrino beam for the NOvA (NuMI Off-axis $\nu_e$ Appearance) long-baseline experiment \cite{NOvA:2019cyt}. Designed to explore fundamental properties of neutrinos, NuMI has provided invaluable data that has contributed to numerous breakthroughs and deepened insights into neutrino interactions and oscillations.

As with any high-intensity accelerator-based system, the successful operation of the NuMI beamline requires careful optimization, monitoring, and maintenance to ensure stable and reliable beam performance. Over the years, the experience gained from NuMI’s operation has highlighted key challenges, such as maintaining beamline stability under varying operational conditions and optimizing performance and evolving experimental demands.

This paper highlights a 
review of the lessons learned from the operation and upkeep of the NuMI beamline, with a focus on the strategies and best practices that have been developed to address these challenges. By sharing these insights, we aim to contribute a roadmap for future projects, including the Long-Baseline Neutrino Facility (LBNF), which will build upon NuMI’s legacy. The experiences and methods discussed here not only enhance our approach to neutrino beamlines but also provide valuable guidelines for future facilities in neutrino research.
\section{Beamline updates}
\begin{figure}[htp]
    \centering
    \includegraphics[width=0.31\linewidth]{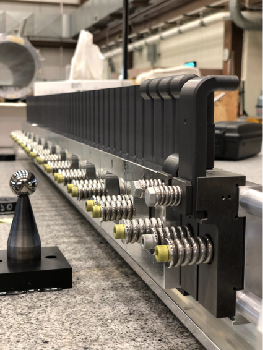}
    \includegraphics[width=0.55\linewidth]{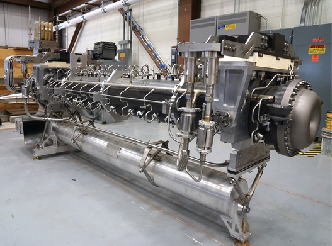}
    \includegraphics[width=0.6\linewidth]{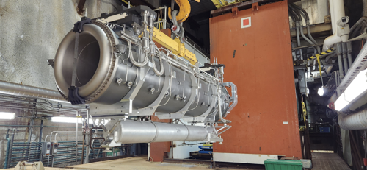}
    \caption{Images of key NuMI beamline components: the 1 MW target (top left), horn 1 (top right), and horn 2 (bottom).}
    \label{fig:1mw_tgt}
\end{figure}

\begin{figure}[htp]
    \centering
    \includegraphics[width=0.7\linewidth]{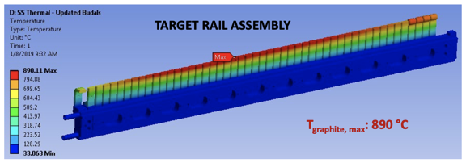}
    \caption{Simulated thermal distribution across the target segments.}
    \label{fig:tgt_temp}
\end{figure}

Over the past few years, key upgrades and maintenance have been conducted to enhance the performance and durability of the NuMI beamline components. In the summer of 2019, the original 700 kW target was replaced with a 1 MW target shown in Fig.~\ref{fig:1mw_tgt} (top left) to support increased beam power. 
The 1 MW target dimensions were modified from the original 700 kW target, with the width increased from 7.4 mm to 9 mm and the height from 14.3 mm to 15.53 mm. These changes were based on detailed mechanical and thermal studies to ensure durability under high-power conditions.
Four cylindrical fins were added to the initial target segments to mitigate thermal shocks on the upstream decay pipe window in accidental events, such as a missteered beam or the beam missing the target.
Dedicated simulation studies were conducted to examine the temperature distribution across the target segments, helping to identify and address critical areas, as illustrated in Fig.~\ref{fig:tgt_temp}.

\begin{figure}[htp]
    \centering
    \includegraphics[width=0.442\linewidth]{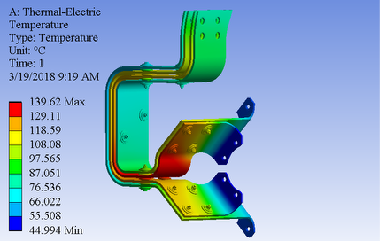}
    \includegraphics[width=0.42\linewidth]{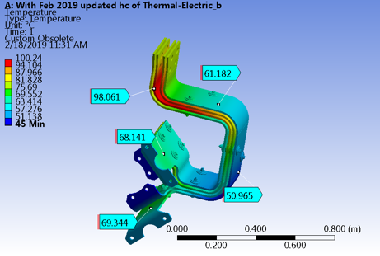}
    \caption{Thermal study results showing the temperature distribution in the stripline system before (left) and after (right) the addition of the air diverter.}
    \label{fig:air_div}
\end{figure}

Following this, in the summer of 2020, both horn 1 and horn 2 were 
replaced to handle 1 MW operations, further preparing the system for higher intensity output. In Fig.~\ref{fig:1mw_tgt} shows the images of prepared horn 1 (top right) and horn 2 (bottom) for 1 MW beam operations.
In the summer of 2022, after reaching the end of the first 1 MW target's service life, the target was replaced to maintain beamline performance and reliability. Most recently, in January 2023, horn 2 was replaced due to a stripline failure as discuss in Sec.~\ref{sec:LL}, ensuring the continuity and stability of operations at the elevated power levels. These upgrades reflect a continuous commitment to advancing the beamline’s capability and longevity.
The 1 MW horn 1 system has been engineered with enhanced cooling capabilities to support high-power operations. A key feature of this system is the addition of an air diverter 
, which provides targeted cooling to the stripline. 
Figure \ref{fig:air_div} shows the temperature distribution across the striplines before (left) and after (right) the air diverter installation on Horn 1.
This air-cooling mechanism has significantly reduced the overall temperature of the stripline, helping to prevent overheating and extend the component's lifespan.
The improved cooling efficiency ensures that the horn system can operate reliably at 1 MW power levels, reducing the risk of temperature-related malfunctions and improving overall system stability.
\begin{figure}
    \centering
    \includegraphics[width=0.3\linewidth]{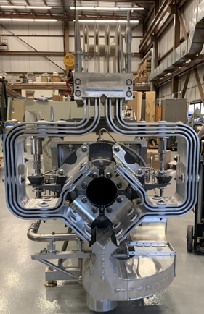}
    \caption{An image of the horn 1 showing the air diverter at the bottom as the upgraded cooling system to the horn striplines.}
    \label{fig:horn_cool}
\end{figure}

\begin{figure}[htp!]
    \centering
    \includegraphics[width=0.5\linewidth]{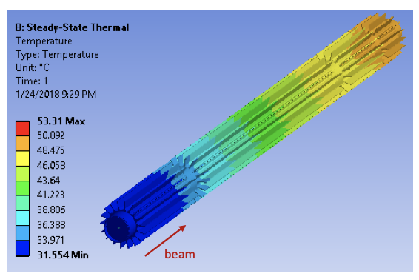}
    \caption{Thermal study results showing the temperature distribution along the baffle.}
    \label{fig:baffle}
\end{figure}

The NuMI beamline baffle is a critical component designed to protect the downstream target and horn systems from potential damage caused by misaligned or errant proton beams.
To support 1 MW operations, the inner diameter of the baffle was increased to accommodate the larger beam spot size on the target, which expanded from 1.2 mm to 1.5 mm. Thermal analysis verified that this adjustment does not affect the baffle's performance.
The increased baffle diameter helps manage heat distribution more effectively, ensuring stable operation and preventing excessive temperature buildup in high-power scenarios. Thermal analysis indicates that under normal operating conditions, the peak temperatures for both the graphite and aluminum components would reach approximately 50 °C as shown in Fig.~\ref{fig:baffle}. 

Beam dispersion in the NuMI beamline was studied by adjusting the lattice parameters of the quadrupole magnets. 
These adjustments optimized the beam optics to enlarge the beam spot size on the target while minimizing dispersion, thereby enhancing beam focus and improving the precision of neutrino delivery.
This optimization process ensures that the beam remains stable and precisely aligned, which is crucial for maintaining consistent experimental conditions.
\section{Lessons Learned}\label{sec:LL}
This section highlights key lessons gained from recent experiences during the beam operations. 
\subsection*{Horn2 Failures}
The NuMI horn striplines are vital components of the horn system, designed to deliver high-current pulses necessary for generating the magnetic fields used to focus charged particles in the neutrino beamline. Each horn is connected to its power supply via four striplines for the NuMI system. These striplines are constructed from high-conductivity aluminum to minimize electrical resistance and efficiently handle the intense current, typically exceeding 200 kA per pulse.
The typical width and thickness of the striplines are approximately 203 mm and 9.5 mm, respectively, ensuring they can withstand the mechanical and thermal stresses associated with high-power operations.

\begin{figure}[htp]
    \centering
    \includegraphics[width=0.23\linewidth]{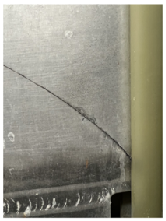}
    \includegraphics[width=0.31\linewidth]{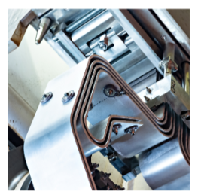}
    \caption{The left image shows the discovered crack on the outermost stripline, extending from the target pile near the endpoint. The right image shows the fracture identified on one of the localized stripline conductors mounted on Horn 2.}
    \label{fig:strip_1}
\end{figure}

At the end of 2023, a stripline failure in the horn system was observed. During inspections, a crack was discovered on the outermost stripline, extending from the target pile near the endpoint directly above the Horn 2 penetration as shown in Fig.~\ref{fig:strip_1} (left). 
Additionally, a fracture was observed in one of the localized stripline conductors mounted on Horn 2 as shown in Fig.~\ref{fig:strip_1} (right). Analysis revealed that these issues were not due to the overall stripline design but were caused by uneven loading on the bent stripline, leading to metal fatigue and eventual failure. A minor defect in the material was also identified, potentially contributing to its weakening and failure. These findings underscore the importance of balanced loading and robust materials to withstand high-power operation demands.
These findings emphasize the importance of load distribution in maintaining the durability of the stripline under high-stress conditions.
\subsection*{Baffle Temperature Issues}
The baffle temperature experiences a notable increase when the beam partially collides or ``scrapes'' against it, as this direct interaction deposits energy into the baffle material. 
Additionally, back-scattered particles from the target contribute to the rise in baffle temperature as shown in Fig~\ref{fig:baffle}. These particles, rebounding from the target area, add further thermal load to the baffle, elevating its temperature beyond normal levels. Together, both beam scraping and back-scattering effects can lead to higher-than-expected baffle temperatures, emphasizing the need for careful beam alignment to manage thermal impacts effectively.
Maintaining a low baffle temperature is essential from both engineering and scientific perspectives. From an engineering standpoint, it prevents the oxidation of the graphite material, ensuring the longevity and reliability of the baffle. Scientifically, reducing neutrino production on the baffle is crucial to minimize uncertainties in the neutrino flux.

\begin{figure}[htp]
    \centering
    \includegraphics[width=0.4\linewidth]{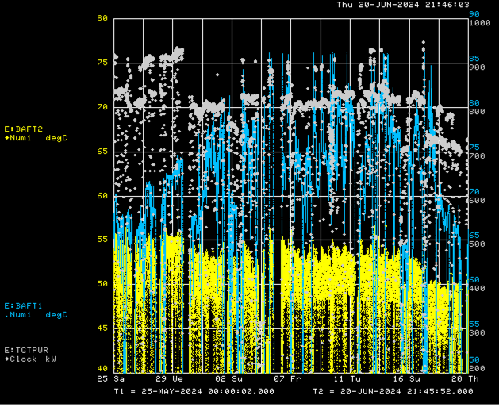}
    \includegraphics[width=0.41\linewidth]{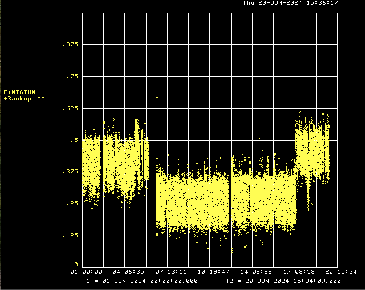}
    \caption{The left image shows the baffle temperature (blue), beam power (gray) and the beam temperature (yellow) during the beam operations from May, 25 to June 20, 2024. The right image shows the horizontal beam position shift at the beam profile monitor ($E:MTGTHM$) in the NuMI beamline.}
    \label{fig:baff_temp_mw}
\end{figure}

Following June 1, 2024, we observed instability in the baffle temperature, which began to push against the beam permit limits. 
Figure \ref{fig:baff_temp_mw} (left) displays the baffle temperature (blue), beam power (gray), and beam temperature (yellow) during operations from May 25 to June 20, 2024. This period also included a shift in beam position, as shown in Fig.~\ref{fig:baff_temp_mw} (right), which we traced to a replacement of the digitizer card on a critical beam position monitor (BPM:TGT) in the NuMI beamline, directly influencing the beam auto-tuning process. 
To address the baffle temperature issue, we optimized the beam position and carefully tuned the Main Injector (MI) chromaticity, reducing the beam spot size impact on the baffle temperature. With these adjustments, we stabilized the baffle temperature and successfully tuned the beam, preparing for the 1 MW operational challenge.
\section{Testing 1 MW challange}
The 1 MW Challenge marks a significant advancement in high-power beam operations, pushing the boundaries of existing technology and infrastructure to maintain stable, high-intensity beam performance. This rigorous testing and optimization process has yielded valuable insights that will guide future high-power accelerator facilities like LBNF.

During testing, the beam power was gradually increased in controlled stages, solidifying the system’s ability to to sustain megawatt-level operations.
As outlined in the previous section, we optimized the beam position and tuned the Main Injector chromaticity to minimize the beam's impact on the baffle temperature.
This tuning and ramp-up process culminated in a full hour of 1 MW beam operation with zero interruptions, marking a significant milestone in beam stability and power handling. Ultimately, the highest recorded average beam power reached 1.018 MW, showcasing the NuMI beamline's robust performance under high-power conditions.
\section{Future targets and horns}
\begin{figure}[htp]
    \centering
    \includegraphics[width=0.9\linewidth]{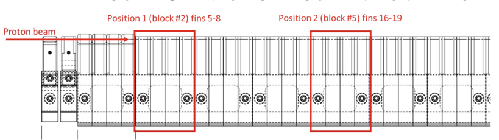}
    \caption{Schematic diagram of the new 1 MW target featuring eight newly designed graphite fins for testing.}
    \label{fig:tgt_new}
\end{figure}

Efforts are underway to enhance the durability and performance of targets and horns to support 1 MW operations, especially for the upcoming Long-Baseline Neutrino Facility (LBNF). A new 1 MW target is in the final stages of preparation, featuring an improved type of graphite material designed to offer superior thermal conductivity. Figure \ref{fig:tgt_new} displays the locations of the eight newly designed graphite segments, highlighted in red boxes within the tightly packed NuMI target. The new graphite materials selected for testing have a slightly higher density than those in the existing target design. 
This advancement aims to better manage the thermal load at high power levels, ensuring reliability and longevity for LBNF targetry. Installation of this new target is scheduled for Fiscal Year 2025.

In parallel, progress is being made on critical components of the beamline’s horn system. The final welds on a spare Horn 1 are nearing completion, with similar work on a spare Horn 2 scheduled to begin in early 2025. These spare horns will provide essential backup support, ensuring that beamline operations remain robust and uninterrupted. Together, these developments reflect a commitment to achieving stable, high-power beam delivery for future experimental needs.
\section{Summary and Outlook}
Significant upgrades and optimizations have been implemented to enable 1 MW operations for the NuMI beamline. These improvements include updates to the target-baffle and horn systems, as well as beam spot size optimization to ensure stable, high-power performance. Key lessons were learned from addressing baffle temperature increases and handling horn component failures, which contributed to enhancing the system's robustness. A successful demonstration of the beamline’s 1 MW capability marked a major milestone. Looking forward, tests of a new graphite composition are planned to further support Long-Baseline Neutrino Facility (LBNF) targetry studies. Additionally, spare horns are being prepared to ensure reliable future beam operations. These efforts collectively strengthen the beamline’s high-power capability and readiness for future experimental demands.

The experiences and lessons learned from the NuMI beamline operations provide a valuable roadmap and practical guidelines for future high-power facilities in neutrino research. 
Managing thermal stresses on critical components like the target, baffle, and horn striplines is a key engineering challenge in high-power beam operations. This demands accurate thermal analysis to identify potential hotspots and the integration of advanced cooling systems to ensure the reliability and longevity of these components under intense operational conditions.
From a scientific perspective, maintaining beam alignment and minimizing uncertainties in neutrino flux through optimized beam optics and baffle design are critical. These insights contribute to a framework for designing reliable, high-efficiency beamline systems capable of sustaining megawatt-level operations. Future facilities, such as the Long Baseline Neutrino Facility (LBNF), can build on these strategies to advance the field of neutrino physics while ensuring operational stability and data precision.
\section{Acknowledgment}
This manuscript has been authored by Fermi Research Alliance, LLC under Contract No. DE-AC02-07CH11359 with the U.S. Department of Energy, Office of Science, Office of High Energy Physics.
\bibliographystyle{elsarticle-num}
\bibliography{mylist} 

\end{document}